\documentclass[lettersize,journal]{IEEEtran}
\usepackage{amsmath,amsfonts}
\usepackage{algorithmic}
\usepackage{algorithm}
\usepackage{array}
\usepackage[caption=false,font=normalsize,labelfont=sf,textfont=sf]{subfig}
\usepackage{textcomp}
\usepackage{stfloats}
\usepackage{url}
\usepackage{verbatim}
\usepackage{graphicx}

\usepackage{cite}
\usepackage{pifont}
\newcommand{\cmark}{\ding{51}}  
\newcommand{\xmark}{\ding{55}}  
\usepackage{multirow} 
\usepackage{array} 
\usepackage{amssymb} 
\usepackage{hyperref} 
\usepackage{acronym}
\usepackage{enumitem}
\usepackage{booktabs}
\usepackage{siunitx}      
\usepackage[most]{tcolorbox}
\usepackage{xcolor}
\usepackage{lipsum}  
\begin{document}

\title{FedP3E: Privacy-Preserving Prototype Exchange for Non-IID IoT Malware Detection in Cross-Silo Federated Learning}


\author{Rami Darwish, Mahmoud Abdelsalam, Sajad Khorsandroo, Kaushik Roy}


\markboth{JOURNAL}%
{Shell \MakeLowercase{\textit{et al.}}: A Sample Article Using IEEEtran.cls for IEEE Journals}


\maketitle

\begin{abstract}
As IoT ecosystems continue to expand across critical sectors, they have become prominent targets for increasingly sophisticated and large-scale malware attacks. The evolving threat landscape, combined with the sensitive nature of IoT-generated data, demands detection frameworks that are both privacy-preserving and resilient to data heterogeneity. Federated Learning (FL) offers a promising solution by enabling decentralized model training without exposing raw data. However, standard FL algorithms such as FedAvg and FedProx often fall short in real-world deployments characterized by class imbalance and non-IID data distributions—particularly in the presence of rare or disjoint malware classes. To address these challenges, we propose FedP3E (Privacy-Preserving Prototype Exchange), a novel FL framework that supports indirect cross-client representation sharing while maintaining data privacy. Each client constructs class-wise prototypes using Gaussian Mixture Models (GMMs), perturbs them with Gaussian noise, and transmits only these compact summaries to the server. The aggregated prototypes are then distributed back to clients and integrated into local training, supported by SMOTE-based augmentation to enhance representation of minority malware classes. Rather than relying solely on parameter averaging, our prototype-driven mechanism enables clients to enrich their local models with complementary structural patterns observed across the federation—without exchanging raw data or gradients. This targeted strategy reduces the adverse impact of statistical heterogeneity with minimal communication overhead. We evaluate FedP3E on the N-BaIoT dataset under realistic cross-silo scenarios with varying degrees of data imbalance. Results demonstrate that FedP3E consistently surpasses FedAvg and FedProx in malware detection performance, achieving accuracy ranging from 95.11\% in severe non-IID conditions to 99.57\% under light heterogeneity.
\end{abstract}



\section{Introduction}
\IEEEPARstart{T}{he} rapid advancement of 5G and the anticipated deployment of Beyond 5G (B5G) technologies are driving the proliferation of IoT devices at an unprecedented scale \cite{rey2022federated}. This hyper-connectivity is enabling new capabilities across healthcare, transportation, industry, and military sectors—collectively referred to as the Extended IoT (XIoT) \cite{darwish2025deep}. However, the expansion of these networks has also introduced a significantly larger attack surface, accelerating both the frequency and severity of cyber threats targeting IoT ecosystems.

As XIoT domains continue to evolve, so do the associated security risks. Real-world incidents have underscored the scale and impact of these vulnerabilities. For instance, large-scale botnets such as Mirai and Gafgyt have been responsible for massive service disruptions and financial damage, with global losses estimated in the billions \cite{website_IoT_malware_loss}. Additionally, the 2018 ransomware attack on the Taiwan Semiconductor Manufacturing Company—the world’s largest chipmaker—exposed the fragility of Industrial IoT (IIoT) systems when operating under insufficiently secured infrastructures \cite{smith2022machine}.

The diverse and heterogeneous nature of XIoT environments makes them especially difficult to protect. IoT devices monitor a wide range of data sources, including sensor readings, system logs, and control commands, often in real time. These data streams may contain highly sensitive or private information. Machine Learning (ML) and Deep Learning (DL) techniques have shown promise in detecting malicious activity across such data sources. However, conventional ML/DL approaches rely on centralized data aggregation, requiring raw data to be transmitted from edge devices to a central server \cite{wu2022adaptive}. This raises serious privacy concerns and regulatory challenges, especially in critical domains \cite{amjath2025graph} like healthcare and smart infrastructure.

Federated Learning (FL) has emerged as a viable alternative to centralized training by enabling decentralized learning directly at the data source. In FL, devices collaboratively train a shared global model without exchanging raw data, thereby preserving user privacy and reducing communication costs. FL is commonly deployed in two paradigms: cross-device \cite{chen2023fs}, where numerous clients (e.g., smartphones) contribute intermittently, and cross-silo \cite{lomurno2022sgde}, where a limited number of reliable organizations (e.g., hospitals, smart factories) participate in regular training rounds. Cross-silo FL is particularly well-suited to IoT malware detection, where devices are often grouped into domains with stable and trusted infrastructures.

Despite its advantages, FL faces several key challenges. A primary issue is the presence of non-independent and identically distributed (non-IID) data across clients. In real-world deployments, the distribution of malware types may vary significantly from one silo to another, leading to local models that converge poorly and global models that underperform on minority classes \cite{darwish2025comparative}. This issue is exacerbated when class imbalance is present—some malware families may be underrepresented or entirely missing on specific clients.

This work is driven by two pressing challenges in federated IoT malware detection: (1) enabling effective knowledge sharing among clients without exposing sensitive local data or requiring overlapping class distributions, and (2) improving generalization for rare or underrepresented malware families, which are frequently neglected by standard FL methods. Existing approaches like FedAvg and FedProx fall short in these scenarios, particularly under non-IID disjoint data conditions. To overcome these limitations, we introduce FedP3E---a novel prototype-based FL framework that transmits class-specific statistical summaries instead of raw data or model gradients. This design preserves data privacy, facilitates learning across heterogeneous silos, and enhances detection performance for minority malware classes, all while maintaining a low communication overhead.

The main contributions of this work are summarized as follows:

\begin{itemize}
\item \textbf{Prototype-Based Representation Sharing:} Instead of exchanging raw data or gradients, each client summarizes its local class-wise distribution using GMM-based prototypes. These representations are perturbed with Gaussian noise to preserve privacy while enabling implicit knowledge sharing across clients.

\item \textbf{Disjoint-Aware Data Augmentation:} We integrate server-aggregated prototypes into local training and apply SMOTE-based oversampling to improve class balance. This addresses the limitations of disjoint and imbalanced data distributions commonly encountered in real-world federated IoT settings.

\item \textbf{Adaptive Communication Mechanism:} Prototype exchange is triggered only when the global model’s accuracy falls below a predefined threshold, ensuring minimal communication overhead while preserving the benefits of collaborative learning.

\item \textbf{Extensive Evaluation on N-BaIoT:} We evaluate FedP3E on the N-BaIoT dataset under both IID and varying degrees of non-IID data imbalance, including disjoint class distributions. The results demonstrate consistent performance improvements over FedAvg and FedProx across heterogeneous federated scenarios.
\end{itemize}

The rest of this paper is organized as follows. Section~\ref{Related work} reviews state-of-the-art research on IoT malware detection using FL, including approaches for handling heterogeneous data and commonly used XIoT malware datasets. Section~\ref{Preliminaries} introduces key terminologies relevant to the FL domain. Section~\ref{Methodological Framework} details the proposed FedP3E framework, including its architecture and underlying techniques. Section~\ref{Experiments and Analysis} presents and analyzes the results of our experimental evaluation. Finally, Section~\ref{Conclusion and Future Work} concludes the paper and outlines potential directions for future research.

\section{Related work}
\label{Related work}
This section presents a comprehensive review of the current state-of-the-art in IoT malware detection using FL, examine advanced strategies designed to address data heterogeneity in FL environments, and summarize the most widely used public datasets that model cyberattacks across various XIoT domains.

\subsection{IoT malware detection using Federated Learning}

FL has recently gained traction as a privacy-preserving paradigm for IoT malware detection. A growing body of research has proposed diverse FL-based frameworks targeting various aspects of malware classification, including robustness, efficiency, class imbalance, and real-world deployment challenges.

Several studies have explored IoT malware detection using FL with real-world datasets. One framework leveraged both supervised and unsupervised FL models—specifically a multi-layer perceptron and autoencoder—trained on the N-BaIoT dataset to detect malware on both seen and unseen IoT devices. This approach demonstrated that federated models could achieve performance comparable to centralized models while maintaining data privacy. The study also evaluated aggregation robustness under adversarial conditions, highlighting vulnerabilities in standard FedAvg and the need for secure aggregation mechanisms \cite{rey2022federated}. Similarly, another study addressed the increasing prevalence of botnet attacks in IoT networks by evaluating the effectiveness of FL in detecting IoT malware traffic while preserving user privacy. Using the N-BaIoT dataset, the authors compared FL-based CNN, LSTM, and GRU models against a centralized baseline. The results showed that FL models, particularly CNN, achieved strong performance in identifying abnormal traffic, further stressing the value of FL in privacy-preserving IoT malware detection \cite{do2024horizontal}. 

In the context of the Industrial Internet of Things (IIoT), FL has also been applied to mitigate privacy concerns and address the heterogeneity across industrial environments. A recent study proposed a federated botnet detection method where multiple industrial enterprises collaboratively train models without sharing raw data. The approach demonstrated high adaptability to local settings and robustness against poisoning attacks \cite{zhou2023federated}. In another effort to enhance scalability and detection performance in FL-based IoT security systems, a study introduced a distributed optimization framework using the Siberian Tiger Optimization (STO) algorithm to fine-tune hyperparameters of a CNN model at the central server. These optimized parameters were then distributed to clients for local training \cite{gupta2025distributed}.

To improve model robustness and label efficiency, FedMalDE introduced a semi-supervised FL framework employing knowledge transfer techniques and a subgraph aggregated capsule network to detect IoT malware \cite{pei2022knowledge}. Similarly, FEDroid targeted Android malware detection using a residual neural network combined with a genetic evolution strategy to simulate and detect malware variants, achieving superior performance across multiple Android datasets \cite{fang2023comprehensive}.

In another line of work, FL was integrated with Markov chains and associative rule learning to classify IoT malware across imbalanced and non-IID datasets. This hybrid approach achieved near-perfect accuracy (99\%) while maintaining runtime performance comparable to centralized methods \cite{d2023privacy}. Complementing this, FED-MAL transformed malware binaries into image representations and used a compact CNN (AM-NET) with adversarial training to enhance generalizability on edge devices \cite{abdel2022efficient}.

One framework focused on minimizing latency and model heterogeneity in FL for networked IoT systems, such as those using Raspberry Pi devices. The authors proposed a cloud-unification method to harmonize on-device models, achieving performance gains between 7\% and 13\%, with only minor increases in training time \cite{shukla2023federated}.

The SIM-FED model combined FL and a lightweight 1D CNN to deliver a robust and privacy-preserving malware detection framework. Evaluated on the IoT-23 dataset, SIM-FED achieved a 99.52\% accuracy and showed resilience to white-box and black-box adversarial attacks, while also reducing computational overhead \cite{nobakht2024sim}.

Another study focused on ransomware detection through decentralized training across multiple clients, demonstrating strong results across all performance metrics. The integration of preprocessing, feature engineering, and collaborative learning led to a scalable and privacy-preserving detection solution \cite{koike2024federated}.

A CNN-based FL model was also proposed to handle imbalanced datasets and intermittent client participation. In this approach, malware binaries were converted into color images to extract visual features, and data augmentation techniques were applied to balance training samples across clients. The system demonstrated strong results in handling both intermittent connectivity and class imbalance \cite{ullah2023privacy}.

Graph-based FL frameworks have also emerged, including Fed-MalGAT, which outperformed its GCN-based counterpart by using multi-head attention for robust classification across federated rounds. Though it introduced additional computational cost, Fed-MalGAT consistently achieved high accuracy, precision, and F1 scores across multiple evaluations \cite{amjath2025graph}.

Recent work has also emphasized control-flow-based static analysis. One study introduced an FL framework for IoT binary classification using CFG-derived features under IID and non-IID scenarios. The models—FL-CNN and FL-DNN—demonstrated strong performance, with FL-CNN achieving 95.27\% accuracy in the IID setting. The study further applied threshold tuning and class weighting to improve results under class imbalance \cite{darwish2025comparative}.

Collectively, these works underscore the growing effectiveness of FL in detecting IoT malware under realistic and constrained conditions.

\subsection{Heterogeneous data processing}

Several advanced strategies have been proposed to address the challenges posed by non-IID data in FL. In \cite{zhang2021client}, a novel algorithm named CSFedAvg was introduced to mitigate accuracy degradation caused by non-IID distributions. This method leverages weight divergence to estimate the degree of data heterogeneity at each client and selects those with more IID-like distributions for more frequent participation, thereby improving global model performance. In \cite{seo2022resource}, a resource-efficient FL framework based on auction theory was proposed to minimize training costs while addressing non-IID effects. The approach jointly optimizes model utility, computational overhead, and data generation expenses, demonstrating that incorporating even a small fraction (less than 1\%) of shared IID data significantly improves training efficiency and stakeholder profitability. 

The FedSLD approach presented in \cite{luo2022fedsld} adjusts the contribution of each local data sample to the training objective based on the client’s label distribution. By addressing label imbalance during optimization, the method improves training stability and convergence in heterogeneous data settings. In \cite{bansal2023fednse}, FedNSE quantifies label distribution skewness and entropy to assess the non-IID characteristics of data across clients. Based on this assessment, it selects an optimal subset of clients for training. Results indicate a reduction of at least 10\% in training loss and improved convergence speed compared to baseline methods.

To address non-IID data from a clustering perspective, \cite{kim2023k} introduced K-FL, a Kalman filter-based FL method that groups clients with similar data distributions to produce low-variance, cluster-specific models. It operates without prior knowledge or initialization settings and achieves faster training and higher accuracy than FedAvg on MNIST, FMNIST, and CIFAR-10 datasets. Lastly, \cite{shu2022clustered} proposed clustered Federated Multitask Learning (FMTL), which integrates model clustering and multitask learning within a dual-server privacy-preserving architecture. Secure two-party computation protocols are used to ensure data privacy while enhancing communication efficiency and overall model performance in non-IID environments.

\subsection{Public datasets in XIoT malware detection}

A wide array of public datasets has enabled significant progress in malware detection across the Extended Internet of Things (XIoT) domains, including IoT, Industrial Internet of Things (IIoT), Internet of Medical Things (IoMT), and Internet of Vehicles (IoV). These datasets vary in attack types, sources, and structural characteristics, supporting diverse detection approaches. Table \ref{datasets} illustrates a summary of public datasets commonly used in XIoT-based malware detection, along with the federated splitting criterion.

To investigate Telnet-based attacks targeting IoT devices, IoTPoT \cite{pa2015iotpot} introduces a honeypot-driven sandbox that captures malware activity across architectures such as ARM, MIPS, and PPC. The dataset highlights the rapid evolution of DDoS malware families and their cross-platform propagation strategies. Bot-IoT \cite{botiot_dataset}, created within a cyber range, simulates realistic network traffic comprising both normal and malicious flows. Covering a broad spectrum of threats—DDoS, keylogging, scanning, and exfiltration—it provides millions of labeled flows for evaluating anomaly-based intrusion detection models. Focusing on protocol-specific traffic from both benign and infected IoT devices, IoT-23 \cite{iot23_dataset} captures 23 real-world scenarios in pcap format. Its curated mix of malware and clean device captures supports studies on behavioral analysis and malware fingerprinting. With a static analysis approach tailored for Android, Drebin \cite{arp2014drebin} is a well-known Android malware dataset introduced alongside a lightweight detection method designed to operate directly on smartphones. Given the rising volume and diversity of malicious Android applications, Drebin addresses the limitations of traditional defenses by employing broad static analysis. It extracts features from various sources—such as permissions, API calls, and network addresses—and embeds them in a joint vector space to capture malware-specific patterns. The dataset contains 123,453 Android applications and 5,560 labeled malware samples. Originating from a Microsoft-hosted classification challenge, BIG 2015 \cite{ronen2018microsoft} features a massive collection of binaries representing nine malware families. Each sample includes both raw hexadecimal content and disassembled metadata, useful for signature- and behavior-based classification. Malimg \cite{nataraj2011malware} takes a visual approach by converting binary files into grayscale images. With over 9,000 samples across 25 families, it allows malware to be classified based on visual texture patterns without requiring disassembly. The Malgenome dataset \cite{zhou2012dissecting} systematizes over 1,200 Android malware samples collected over a year. It provides deep insights into mobile malware evolution, installation methods, and evasion tactics. Improving upon KDD’99, NSL-KDD \cite{tavallaee2009detailed} removes redundancy and balances class distribution. It remains a foundational benchmark for intrusion detection systems with well-defined categories like DoS, U2R, R2L, and probing. Designed for medical environments, WUSTL-EHMS-2020 \cite{hady2020intrusion} integrates biometric sensor data with network flow metrics. It captures spoofing and data injection attacks in a real-time healthcare testbed, reflecting IoMT-specific threats. Finally, N-BaIoT \cite{meidan2018n} offers traffic data from nine IoT devices infected with Mirai and BASHLITE. By using deep autoencoders to model benign behavior, the dataset facilitates precise detection of device-level anomalies in dynamic enterprise settings.

\begin{table}[!t]
\caption{Summary of Public Datasets Commonly Used in XIoT Malware Detection}
\label{datasets}
\centering
\footnotesize
\setlength{\tabcolsep}{2.5pt}
\begin{tabular}{|p{3.2cm}|p{1.7cm}|p{3.5cm}|}
\hline
\textbf{Dataset} & \textbf{XIoT Domain} & \textbf{Federated Splitting Criterion} \\
\hline
IoT-23 \cite{garcia2020iot} & IoT & PCAP \\
\hline
N-BaIoT \cite{meidan2018n} & IoT & IoT device \\
\hline
IoTPoT \cite{pa2015iotpot} & IoT & Botnet sample / traffic stream \\
\hline
Bot-IoT \cite{koroniotis2019towards} & IIoT & Attack type \\
\hline
BIG 2015 \cite{ronen2018microsoft} & IoT/IIoT/IoMT & Family class \\
\hline
WUSTL-EHMS-2020 \cite{hady2020intrusion} & IoMT & Not applicable \\
\hline
NSL-KDD \cite{tavallaee2009detailed} & IoMT & Traffic category / attack type \\
\hline
Drebin \cite{arp2014drebin} & IoT/IIoT/IoV & Android malware family \\
\hline
Malgenome \cite{zhou2012dissecting} & IoT & Malware type \\
\hline
Malimg \cite{nataraj2011malware} & IIoT/IoV & Malware family \\
\hline
\end{tabular}
\end{table}

\section{Preliminaries}
\label{Preliminaries}
This section provides an overview of key terminologies commonly used in the FL domain, including FL deployment settings, data distribution strategies, and baseline aggregation algorithms.

\subsection{Federated Learning deployment settings}
\subsubsection{Cross-Device}
Cross-device FL refers to a setting in which numerous heterogeneous edge devices—such as smartphones, sensors, and embedded systems—collaboratively train a shared model while keeping their data locally. This setting introduces several system-level challenges that must be addressed to ensure effective deployment \cite{chen2023fs}:

\begin{itemize}
    \item \textbf{Usability and Efficiency:} Participating devices often vary significantly in hardware (e.g., x86, ARM) and software environments. These devices typically have limited computational power, storage capacity, and communication bandwidth, necessitating lightweight and efficient training, inference, and model management procedures.
    
    \item \textbf{Scalability and Robustness:} The system must support a vast number of devices and remain robust in the face of unreliable participants, including those that may frequently disconnect or respond slowly. As more resources are added, the system should scale efficiently to enhance training speed and model quality.
    
    \item \textbf{Flexibility and Extensibility:} Given the diversity of software stacks and APIs across devices, FL frameworks must allow for flexible algorithm customization and extension to optimize convergence and model performance across various application scenarios.
\end{itemize}

\subsubsection{Cross-Silo}

Cross-silo FL refers to collaborative model training among a small number of reliable and resource-rich organizations—such as hospitals, banks, and research institutions—where each participant holds large volumes of private data and remains involved throughout the training process \cite{huang2022cross}. Unlike cross-device FL, cross-silo FL faces unique challenges stemming from the need to balance high model performance, strict privacy requirements, and long-term cooperation between strategically motivated clients. A key technical challenge is statistical heterogeneity, where non-IID data distributions across clients degrade global model performance; this is addressed through data moderation, client clustering, and personalization techniques, which are more privacy-preserving. While system heterogeneity is less concerning due to clients' robust infrastructure, optimizing communication and computation through model compression and selective client participation remains important. Privacy and security are paramount, with defenses including differential privacy, homomorphic encryption, and secure multi-party computation—all offering tradeoffs between protection strength and computational cost. Moreover, cooperation and incentive mechanisms such as data valuation, profit allocation, and fairness-aware strategies are essential for sustaining client participation \cite{huang2022cross}.

\subsection{Federated Learning data distribution strategies}
\subsubsection{IID data in FL} In FL, data is considered independent and identically distributed (IID) when each client’s local dataset follows the same underlying probability distribution and consists of statistically similar and independent samples. This assumption simplifies the collaborative training process by ensuring that all clients contribute uniformly representative data. Under IID conditions, local model updates are more aligned, reducing inter-client variability and enabling faster and more stable convergence of the global model. Consequently, performance discrepancies between local and global models are minimal, limiting the need for additional techniques such as personalization. Many foundational FL algorithms, such as FedAvg, were initially designed under this idealized setting, although real-world applications often deviate from these assumptions. The IID scenario is typically used as a performance baseline when evaluating FL algorithms and protocols in both simulations and empirical studies. It offers a controlled environment to understand the core behavior of FL frameworks before addressing more realistic data challenges such as statistical heterogeneity and privacy risks \cite{lu2024federated}.

\subsubsection{Non-IID data in FL} In practice, FL often operates under non-IID conditions, where data distributions vary significantly across clients. This heterogeneity stems from differences in user behavior, application usage, geographic location, sensor configurations, data collection frequency, and demographic characteristics, leading to inconsistencies in data quantity, class representation, and feature distributions \cite{lu2024federated}. These disparities introduce critical challenges such as increased communication costs, slower model convergence, reduced generalization performance, and heightened risks of privacy leakage. Non-uniform data distributions across clients require more frequent communication with the server to reach model consensus, resulting in higher latency and less efficient training. Class imbalance across clients can cause local models to underperform on underrepresented categories, which negatively affects the global model. Moreover, conflicting local gradients can hinder convergence, and feature-based discrepancies may reveal sensitive client information during aggregation \cite{altomare2024client}. 

\subsection{Baseline optimization and aggregation strategies in Federated Learning}
FedAvg (Federated Averaging) \cite{mcmahan2017communication}  and FedProx (Federated Proximal) \cite{li2020federated} are foundational optimization strategies in FL, designed to train models across decentralized data sources. FedAvg operates by having each client perform a fixed number of local gradient descent steps on their private data. Once completed, clients send their updated model parameters to a central server, which averages these updates—weighted by the relative size of each client's dataset—to form a new global model. This method is simple and communication-efficient, but it does not explicitly restrict how far local updates can drift from the global model, which can become problematic in the presence of non-IID data \cite{su2021non}.

In contrast, FedProx modifies the local objective function at each client by adding a proximal term that penalizes large deviations from the current global model. This means each client not only minimizes its local empirical loss but also includes a regularization term that discourages straying too far from the global model parameters received at the beginning of the round. As a result, FedProx helps stabilize the training process when client data distributions are different. While both algorithms use a similar aggregation approach—computing a weighted average of the client models—the key distinction lies in how local models are updated: FedAvg relies purely on local descent, while FedProx constrains the updates to remain close to the shared model. This difference makes FedProx particularly effective in handling heterogeneous data environments where client updates may diverge significantly \cite{su2021non}.

\section{Methodological Framework}
\label{Methodological Framework}
This section presents the complete pipeline and techniques employed in the proposed FedP3E framework. The procedural steps are outlined in Algorithm~\ref{alg:P3E}.

\subsection{Federated Learning setup}
\begin{figure}[!t]
\centering
\includegraphics[width=\columnwidth]{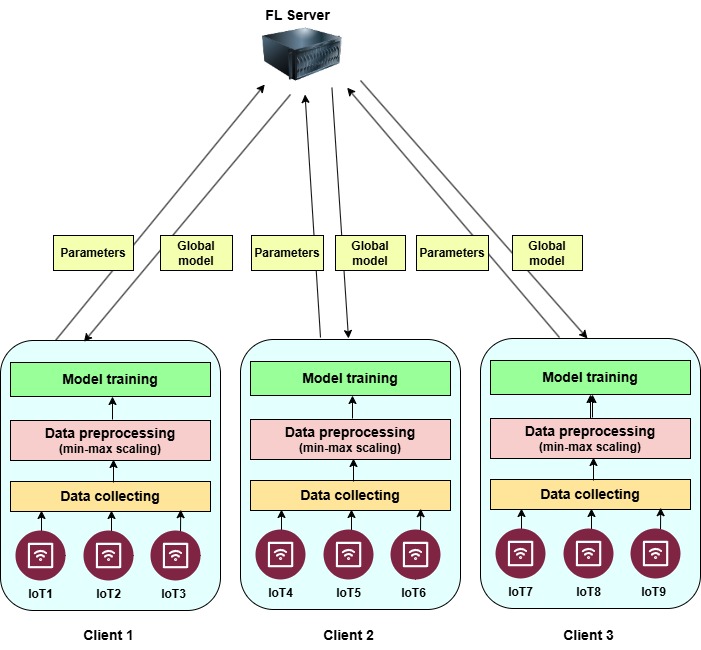}
\caption{Cross-silo federated learning architecture with three client devices collaborating under a central server.}
\label{cross-silo1}
\end{figure}

We employ a FL architecture comprising a central server and three client devices. Each client is a powerful and reliable machine capable of collecting traffic from IoT devices within the same network, for example, operating as a B5G base station—an arrangement that aligns with the cross-silo FL paradigm \cite{rey2022federated}. Each client (silo) acquires data from three distinct IoT devices, performs preprocessing, and conducts local model training, as illustrated in Fig.~\ref{cross-silo1}. The training process spans 20 communication rounds, with each client training locally for 15 epochs per round and transmitting updated model parameters to the server for aggregation.

At the beginning of each round, the server distributes the global model to all clients. If the average accuracy of the global model at round~\(N\) falls below a predefined threshold (which may vary depending on the specific objectives of the application), clients are prompted to generate one-time, class-wise feature prototypes using Gaussian Mixture Models (GMMs). These prototypes, which capture intra-class variability, are transmitted to the server, aggregated, and redistributed to assist clients with missing or underrepresented classes.

To evaluate the proposed framework, we implement two widely adopted FL baselines: FedAvg and FedProx. FedAvg aggregates client updates through weighted averaging, while FedProx incorporates a proximal term into the local objective to mitigate update divergence in non-IID data settings.

\subsection{Dataset description}

To evaluate the proposed FedP3E framework, we utilize the N-BaIoT \cite{meidan2018n} dataset, which comprises network traffic captured from nine distinct IoT devices under both benign and malicious conditions, as summarized in Table~\ref{nbaiot_summary}. Malicious activity was induced by infecting the devices with two malware families: Mirai and BASHLITE (also known as Gafgyt). Each data instance corresponds to a network packet recorded using Wireshark and is represented by 115 numerical features that reflect statistical properties—such as packet size, count, and jitter—computed over various time windows (ranging from 100 milliseconds to 1 minute).

The dataset is structured such that each device maintains a separate log, which facilitates natural partitioning for FL experiments. All devices include both benign and Gafgyt-infected traffic, while seven out of the nine devices contain Mirai-related samples. Specifically, the Ennio doorbell and the Samsung SNH 1011 N webcam do not exhibit Mirai infections. Table~\ref{Malware_distribution} details the distribution of Gafgyt and Mirai variants across all IoT devices.

\paragraph*{Why N-BaIoT?}
N-BaIoT dataset is particularly suited to our study for three reasons:
\begin{enumerate}
    \item \textbf{Cross-silo mapping} — The per-device log structure naturally maps to a cross-silo FL setting in which each silo aggregates traffic from one or more IoT devices managed by the same edge node.
    \item \textbf{Multi-class challenge} — The coexistence of benign traffic and multiple malware families yields a challenging, realistic multi-class detection task.
    \item \textbf{Configurable non-IID} — The dataset structure allows for simulating varying degrees of non-IID conditions, including: label distribution skew, where clients observe different class proportions; 
quantity skew, where the number of samples per client varies significantly; and disjoint distribution, where clients possess mutually exclusive sets of class labels.
\end{enumerate}

This flexibility is essential for systematically validating the robustness of FedP3E under varying degrees of data heterogeneity.

\subsection{Client data characteristics and data distribution}

To emulate realistic cross-silo FL deployments, we consider two primary data distribution strategies across clients: \textit{IID} and \textit{non-IID} scenarios.

In the IID setting, the dataset is partitioned evenly among three clients based on device types. Specifically, all traffic data from IoT devices 1, 2, and 3 are assigned to Client 1; devices 4, 5, and 6 to Client 2; and the remaining devices (7, 8, and 9) to Client 3. This setup ensures that each client receives a balanced mix of benign samples, Gafgyt variants, and Mirai variants, thereby maintaining uniform class representation across the federation, as illustrated in Table~\ref{IID_distribution}.

In contrast, the non-IID distribution is explored through three escalating levels of statistical heterogeneity:

\begin{itemize}
    \item \textbf{Case 1: Light non-IID} — Certain attack variants are more prevalent in one client while sparsely present or entirely absent in others. For instance, the Gafgyt variant \textit{combo} is abundant in Client 1, lightly represented in Client 3, and completely absent in Client 2, as shown in Table~\ref{Non-IID_light_distribution}.
    
    \item \textbf{Case 2: Moderate non-IID} — Some attack types are entirely missing from one or more clients. For example, the Mirai variant \textit{udp plain} appears exclusively in Client 2 and is absent from both Clients 1 and 3, as illustrated in Table~\ref{Non-IID_Moderate_distribution}.
    
    \item \textbf{Case 3: Severe non-IID} — This most challenging configuration introduces extreme class imbalance where each client is assigned data from only one category. Specifically, Client 1 contains only benign samples, Client 2 hosts only Gafgyt samples, and Client 3 is limited to Mirai variants. This severe partitioning simulates a highly heterogeneous federated environment, as illustrated in Table~\ref{Non-IID_Severe_distribution}, and poses significant challenges for model convergence and generalization due to the complete absence of overlapping classes across clients.
\end{itemize}

These non-IID scenarios introduce practical challenges commonly encountered in distributed IoT environments. Differences in device capabilities, usage patterns, and threat exposure often result in skewed or incomplete local data distributions. As such, our design reflects real-world heterogeneity, where local models may struggle to learn from limited or biased representations of the overall data distribution.

\begin{algorithm}[H]
\caption{FedP3E: Federated Privacy-Preserving Prototype Exchange}
\label{alg:P3E}
\begin{algorithmic}[1]
\REQUIRE Federated clients $\{C_1, C_2, ..., C_K\}$, global rounds $T$, prototype exchange threshold $\tau$, evaluation round $r^*$
\ENSURE Trained global model $\mathcal{M}_G$

\STATE Initialize global model $\mathcal{M}_G^0$
\FOR{each round $t = 1$ to $T$}
    \STATE Server sends $\mathcal{M}_G^{t-1}$ to all clients
    \FOR{each client $C_k$ in parallel}
        \STATE Perform local training on $C_k$ for $E$ epochs using local data $\mathcal{D}_k$
        \STATE Send updated model $\mathcal{M}_k^t$ to the server
    \ENDFOR
    \STATE Server aggregates $\{\mathcal{M}_k^t\}_{k=1}^K$ to update $\mathcal{M}_G^t$
    
    \IF{$t = r^*$ \AND Accuracy$(\mathcal{M}_G^t) < \tau$}
        \FOR{each client $C_k$ in parallel}
            \FOR{each class $c \in \mathcal{C}_k$}
                \STATE Fit a GMM to local features $\mathcal{X}_c$
                \STATE Extract component means $\mu_j^{(c)}$ as prototypes
                \STATE Perturb each prototype: $\tilde{\mu}_j^{(c)} = \mu_j^{(c)} + \epsilon, \; \epsilon \sim \mathcal{N}(0, \sigma^2 I)$
            \ENDFOR
            \STATE Send $\{\tilde{\mu}_j^{(c)}\}$ to server
        \ENDFOR
        
        \STATE Server aggregates all prototypes by class using MiniBatch K-Means
        \STATE Server sends merged global prototypes $\{\hat{\mu}_l^{(c)}\}$ to all clients
        
        \FOR{each client $C_k$ in parallel}
            \STATE Apply SMOTE to $\{\hat{\mu}_l^{(c)}\}$ to generate synthetic samples
            \STATE Add synthetic data to local dataset $\mathcal{D}_k$
        \ENDFOR
    \ENDIF
\ENDFOR
\RETURN Final global model $\mathcal{M}_G^T$
\end{algorithmic}
\end{algorithm}

\begin{table}[!t]
\caption{Summary of Benign and Attack Instances for Each IoT Device in the N-BaIoT Dataset, Including Infection Status by Gafgyt and Mirai Malware \cite{meidan2018n}}
\label{nbaiot_summary}
\centering
\begin{tabular}{|p{3.7cm}|p{0.7cm}|p{0.6cm}|p{0.8cm}|p{0.9cm}|}
\hline
\textbf{IoT Device} & \textbf{Gafgyt} & \textbf{Mirai} & \textbf{Benign} & \textbf{Attacks} \\
\hline
Danmini Doorbell & \cmark & \cmark & 49,548 & 968,750 \\
\hline
Ennio Doorbell & \cmark & \xmark & 39,100 & 316,400 \\
\hline
Ecobee Thermostat & \cmark & \cmark & 13,113 & 822,763 \\
\hline
Philips B120N/10 Baby Monitor & \cmark & \cmark & 175,240 & 923,437 \\
\hline
Provision PT-737E Security Cam & \cmark & \cmark & 62,154 & 766,106 \\
\hline
Provision PT-838 Security Cam & \cmark & \cmark & 98,514 & 738,377 \\
\hline
SimpleHome XCS7-1002-WHT Cam & \cmark & \cmark & 46,585 & 816,471 \\
\hline
SimpleHome XCS7-1003-WHT Cam & \cmark & \cmark & 19,528 & 831,298 \\
\hline
Samsung SNH-1011N Webcam & \cmark & \xmark & 52,150 & 323,072 \\
\hline
\end{tabular}
\end{table}

\begin{table*}[t]
\centering
\caption{Distribution of Gafgyt and Mirai attack variants across IoT devices in the N-BaIoT dataset}
\begin{tabular}{|p{5.3cm}|p{0.8cm}||p{0.7cm}|p{0.4cm}|p{0.4cm}|p{0.4cm}|p{0.4cm}||p{0.4cm}|p{0.4cm}|p{0.4cm}|p{0.4cm}|p{1.1cm}|}
\hline
\textbf{IoT device} & \textbf{Benign} & \multicolumn{5}{c||}{\textbf{Gafgyt variants}} & \multicolumn{5}{c|}{\textbf{Mirai variants}} \\
\hline

IoT1 (Danmini Doorbell) & Benign & combo & junk & scan & tcp & udp & ack & scan & syn & udp & udp plain \\
\hline
IoT2 (Ecobee Thermostat) & Benign & combo & junk & scan & tcp & udp & ack & scan & syn & udp & udp plain \\
\hline
IoT3 (Ennio Doorbell) & Benign & combo & junk & scan & tcp & udp & \xmark & \xmark & \xmark & \xmark & \xmark \\
\hline
IoT4 (Philips B120N/10 Baby Monitor) & Benign & combo & junk & scan & tcp & udp & ack & scan & syn & udp & udp plain \\
\hline
IoT5 (Provision PT-737E Security Cam) & Benign & combo & junk & scan & tcp & udp & ack & scan & syn & udp & udp plain \\
\hline
IoT6 (Provision PT-838 Security Cam) & Benign & combo & junk & scan & tcp & udp & ack & scan & syn & udp & udp plain \\
\hline
IoT7 (Samsung SNH-1011N Webcam) & Benign & combo & junk & scan & tcp & udp & \xmark & \xmark & \xmark & \xmark & \xmark \\
\hline
IoT8 (SimpleHome XCS7-1002-WHT Cam) & Benign & combo & junk & scan & tcp & udp & ack & scan & syn & udp & udp plain \\
\hline
IoT9 (SimpleHome XCS7-1003-WHT Cam)  & Benign & combo & junk & scan & tcp & udp & ack & scan & syn & udp & udp plain \\
\hline
\end{tabular}
\label{Malware_distribution}
\end{table*}

\begin{table*}[t]
\centering
\caption{Scenario~1: IID distribution of IoT traffic among clients.  
Each client receives a balanced, representative subset of benign packets and every Gafgyt / Mirai variant, emulating an ideal federated learning environment.  
The numbers reported in the table denote the exact instance counts for benign traffic and each malware family}
\begin{tabular}{|p{0.7cm}|p{0.4cm}|p{1cm}||p{0.85cm}|p{0.85cm}|p{0.85cm}|p{1cm}|p{1cm}||p{1cm}|p{1cm}|p{1cm}|p{1cm}|p{1cm}|p{1cm}|}
\hline
\textbf{Client} & \textbf{IoT} & \textbf{Benign} & \multicolumn{5}{c||}{\textbf{Gafgyt variants}} & \multicolumn{5}{c|}{\textbf{Mirai variants}} \\
\cline{4-8} \cline{9-13}
 &  &  & \textbf{Combo} & \textbf{Junk} & \textbf{Scan} & \textbf{Tcp} & \textbf{Udp} & \textbf{Ack} & \textbf{Scan} & \textbf{Syn} & \textbf{Udp} & \textbf{Udp plain} \\
\hline

\multirow{3}{*}{Client1} & IoT1 &  \cmark 49,548 & \cmark 59,718 & \cmark 29,068 & \cmark 29,849 & \cmark 92,141 & \cmark 105,874 & \cmark 102,195 & \cmark 107,685 & \cmark 122,573 & \cmark 237,665 & \cmark 81,982 \\
 & IoT2 &  \cmark 13,113 & \cmark 53,012 & \cmark 30,312 & \cmark 27,494 & \cmark 95,021 & \cmark 104,791 & \cmark 113,285 & \cmark 43,192 & \cmark 116,807 & \cmark 151,481 & \cmark 87,368 \\
 & IoT3 &  \cmark 39,100 & \cmark 53,014 & \cmark 29,797 & \cmark 28,120 & \cmark 101,536 & \cmark 103,933 & \xmark & \xmark & \xmark & \xmark & \xmark \\
\hline
\multirow{3}{*}{Client2} & IoT4 &  \cmark 175,240 & \cmark 58,152 & \cmark 28,349 & \cmark 27,859 & \cmark 92,581 & \cmark 105,782 & \cmark 91,123 & \cmark 103,621 & \cmark 118,128 & \cmark 217,034 & \cmark 80,808 \\
 & IoT5 & \cmark 62,154 & \cmark 61,380 & \cmark 30,898 & \cmark 29,297 & \cmark 104,510 & \cmark 104,510 & \cmark 60,554 & \cmark 96,781 & \cmark 65,746 & \cmark 156,248 & \cmark 56,681 \\
 & IoT6 &\cmark 98,514 & \cmark 57,530 & \cmark 29,068 & \cmark 28,397 & \cmark 89,387 & \cmark 104,658 & \cmark 57,997 & \cmark 97,096 & \cmark 61,851 & \cmark 158,608 & \cmark 53,785 \\
\hline
\multirow{3}{*}{Client3} & IoT7 & \cmark 52,150 & \cmark 58,669 & \cmark 28,305 & \cmark 27,698 & \cmark 97,783 & \cmark 110,617 & \xmark & \xmark & \xmark & \xmark & \xmark \\
 & IoT8 &  \cmark 46,585 & \cmark 54,283 & \cmark 28,579 & \cmark 27,825 & \cmark 88,816 & \cmark 103,720 & \cmark 111,480 & \cmark 45,930 & \cmark 125,715 & \cmark 151,879 & \cmark 78,244 \\
 & IoT9 &  \cmark 19,528 & \cmark 59,398 & \cmark 27,413 & \cmark 28,572 & \cmark 98,075 & \cmark 102,980 & \cmark 107,187 & \cmark 43,674 & \cmark 122,479 & \cmark 157,084 & \cmark 84,436 \\
\hline
\end{tabular}
\label{IID_distribution}
\end{table*}

\begin{table*}[t]
\centering
\caption{Scenario~2 – Case~1: Light non-IID distribution of attack variants across clients.  
This setting introduces mild statistical heterogeneity, where certain malware variants are unevenly distributed among clients.  
The numbers shown in the table indicate the number of instances per client for both benign samples and specific malware variants}
\begin{tabular}{|p{0.7cm}|p{0.4cm}|p{1cm}||p{0.85cm}|p{0.85cm}|p{0.85cm}|p{1cm}|p{1cm}||p{1cm}|p{1cm}|p{1cm}|p{1cm}|p{1cm}|p{1cm}|}
\hline
\textbf{Client} & \textbf{IoT} & \textbf{Benign} & \multicolumn{5}{c||}{\textbf{Gafgyt variants}} & \multicolumn{5}{c|}{\textbf{Mirai variants}} \\
\cline{4-8} \cline{9-13}
 &  &  & \textbf{Combo} & \textbf{Junk} & \textbf{Scan} & \textbf{Tcp} & \textbf{Udp} & \textbf{Ack} & \textbf{Scan} & \textbf{Syn} & \textbf{Udp} & \textbf{Udp plain} \\
\hline

\multirow{3}{*}{Client1} & IoT1 &  \cmark 49,548 & \cmark 59,718 & \xmark & \xmark  & \cmark 92,141 & \xmark  & \xmark  & \cmark 107,685 & \xmark  & \cmark 237,665 & \cmark 81,982 \\
 & IoT2 &  \cmark 13,113 & \cmark 53,012 & \xmark  & \xmark & \cmark 95,021 & \xmark & \xmark  & \cmark 43,192 & \cmark 116,807 & \xmark  & \cmark 87,368 \\
 & IoT3 &  \cmark 39,100 & \cmark 53,014 & \xmark & \xmark & \cmark 101,536 & \xmark  & \xmark & \xmark & \xmark & \xmark & \xmark \\
\hline
\multirow{3}{*}{Client2} & IoT4 &  \cmark 175,240 & \xmark & \cmark 28,349 & \cmark 27,859 & \xmark  & \cmark 105,782 & \xmark  & \cmark 103,621 & \cmark 118,128 & \cmark 217,034 & \xmark \\
 & IoT5 & \cmark 62,154 & \xmark  & \cmark 30,898 & \cmark 29,297 & \xmark  & \xmark  & \xmark & \xmark  & \cmark 65,746 & \xmark & \xmark \\
 & IoT6 &\cmark 98,514 & \xmark  & \cmark 29,068 & \cmark 28,397 & \xmark  & \xmark  & \xmark  & \xmark  & \cmark 61,851 & \xmark  & \xmark  \\
\hline
\multirow{3}{*}{Client3} & IoT7 & \cmark 52,150 & \cmark 58,669 & \xmark  & \xmark  & \cmark 97,783 & \xmark  & \xmark & \xmark & \xmark & \xmark & \xmark \\
 & IoT8 &  \cmark 46,585 & \xmark  & \xmark  & \xmark  & \xmark  & \cmark 103,720 & \cmark 111,480 & \xmark  & \xmark  & \xmark  & \cmark 78,244 \\
 & IoT9 &  \cmark 19,528 & \xmark  & \cmark 27,413 & \xmark  & \xmark  & \cmark 102,980 & \cmark 107,187 & \xmark  & \xmark  & \cmark 157,084 & \cmark 84,436 \\
\hline
\end{tabular}
\label{Non-IID_light_distribution}
\end{table*}

\begin{table*}[t]
\centering
\caption{Scenario~2 – Case~2: Moderate non-IID distribution of attack variants across clients.  
This setting introduces a higher level of statistical heterogeneity compared to Case~1. Certain malware variants are entirely absent from one or more clients, creating noticeable class imbalance.  
The numbers presented in the table reflect the instance counts of benign and malware samples distributed across clients}
\begin{tabular}{|p{0.7cm}|p{0.4cm}|p{1cm}||p{0.85cm}|p{0.85cm}|p{0.85cm}|p{1cm}|p{1cm}||p{1cm}|p{1cm}|p{1cm}|p{1cm}|p{1cm}|p{1cm}|}
\hline
\textbf{Client} & \textbf{IoT} & \textbf{Benign} & \multicolumn{5}{c||}{\textbf{Gafgyt variants}} & \multicolumn{5}{c|}{\textbf{Mirai variants}} \\
\cline{4-8} \cline{9-13}
 &  &  & \textbf{Combo} & \textbf{Junk} & \textbf{Scan} & \textbf{Tcp} & \textbf{Udp} & \textbf{Ack} & \textbf{Scan} & \textbf{Syn} & \textbf{Udp} & \textbf{Udp plain} \\
\hline

\multirow{3}{*}{Client1} & IoT1 &  \cmark 49,548 & \cmark 59,718 & \xmark & \xmark & \xmark & \cmark 105,874 & \cmark 102,195 & \xmark & \cmark 122,573 & \xmark & \xmark \\
 & IoT2 &  \cmark 13,113 & \cmark 53,012 & \xmark & \xmark & \xmark & \xmark & \cmark 113,285 & \xmark & \cmark 116,807 & \xmark & \xmark \\
 & IoT3 &  \cmark 39,100 & \cmark 53,014 & \xmark & \xmark & \xmark & \xmark & \xmark & \xmark & \xmark & \xmark & \xmark \\
\hline
\multirow{3}{*}{Client2} & IoT4 &  \cmark 175,240 & \xmark & \cmark 28,349 & \xmark & \cmark 92,581 & \xmark & \xmark & \cmark 103,621 & \cmark 118,128 & \xmark & \cmark 80,808 \\
 & IoT5 & \xmark & \xmark & \cmark 30,898 & \xmark & \cmark 104,510 & \cmark 104,510 & \xmark & \cmark 96,781 & \cmark 65,746 & \xmark & \cmark 56,681 \\
 & IoT6 &\xmark & \xmark & \cmark 29,068 & \xmark & \cmark 89,387 & \xmark & \xmark & \cmark 97,096 & \cmark 61,851 & \xmark & \cmark 53,785 \\
\hline
\multirow{3}{*}{Client3} & IoT7 &\xmark & \xmark & \xmark & \cmark 27,698 & \xmark & \xmark & \xmark & \xmark & \xmark & \xmark & \xmark \\
 & IoT8 & \xmark & \xmark & \xmark & \cmark 27,825 & \xmark & \xmark & \cmark 111,480 & \xmark & \xmark & \cmark 151,879 & \xmark \\
 & IoT9 &  \cmark 19,528 & \xmark & \xmark & \cmark 28,572 & \xmark & \cmark 102,980 & \cmark 107,187 & \xmark & \xmark & \cmark 157,084 & \xmark \\
\hline
\end{tabular}
\label{Non-IID_Moderate_distribution}
\end{table*}

\begin{table*}[t]
\centering
\caption{Scenario~2 – Case~3: Severe non-IID distribution of attack variants across clients.  
In this setting, each client is exposed to a disjoint subset of the class space: Client~1 contains only benign samples, Client~2 observes exclusively Gafgyt malware, and Client~3 is limited to Mirai malware.  
The numbers reported in the table indicate the instance counts for each class type (benign, Gafgyt, and Mirai) assigned to the respective clients}

\begin{tabular}{|p{0.7cm}|p{0.4cm}|p{1cm}||p{0.85cm}|p{0.85cm}|p{0.85cm}|p{1cm}|p{1cm}||p{1cm}|p{1cm}|p{1cm}|p{1cm}|p{1cm}|p{1cm}|}
\hline
\textbf{Client} & \textbf{IoT} & \textbf{Benign} & \multicolumn{5}{c||}{\textbf{Gafgyt variants}} & \multicolumn{5}{c|}{\textbf{Mirai variants}} \\
\cline{4-8} \cline{9-13}
 &  &  & \textbf{Combo} & \textbf{Junk} & \textbf{Scan} & \textbf{Tcp} & \textbf{Udp} & \textbf{Ack} & \textbf{Scan} & \textbf{Syn} & \textbf{Udp} & \textbf{Udp plain} \\
\hline

\multirow{3}{*}{Client1} & IoT1 &  \cmark 49,548 & \xmark &\xmark &\xmark & \xmark & \xmark & \xmark & \xmark & \xmark & \xmark & \xmark \\
 & IoT2 &  \cmark 13,113 & \xmark &\xmark &\xmark & \xmark & \xmark & \xmark & \xmark & \xmark & \xmark & \xmark \\
 & IoT3 &  \cmark 39,100 & \xmark &\xmark &\xmark & \xmark & \xmark & \xmark & \xmark & \xmark & \xmark & \xmark \\
\hline
\multirow{3}{*}{Client2} & IoT4 &  \xmark & \cmark 58,152 & \cmark 28,349 & \cmark 27,859 & \cmark 92,581 & \cmark 105,782 & \xmark & \xmark & \xmark & \xmark & \xmark \\
 & IoT5 & \xmark & \cmark 61,380 & \cmark 30,898 & \cmark 29,297 & \cmark 104,510 & \cmark 104,510 & \xmark & \xmark & \xmark & \xmark & \xmark \\
 & IoT6 & \xmark & \cmark 57,530 & \cmark 29,068 & \cmark 28,397 & \cmark 89,387 & \cmark 104,658 & \xmark & \xmark & \xmark & \xmark & \xmark \\
\hline
\multirow{3}{*}{Client3} & IoT7 & \xmark & \xmark &\xmark & \xmark & \xmark & \xmark & \xmark & \xmark & \xmark & \xmark & \xmark \\
 & IoT8 &  \xmark &  \xmark & \xmark &\xmark & \xmark & \xmark &  \cmark 111,480 & \cmark 45,930 & \cmark 125,715 & \cmark 151,879 & \cmark 78,244 \\
 & IoT9 & \xmark &  \xmark & \xmark &\xmark & \xmark & \xmark &  \cmark 107,187 & \cmark 43,674 & \cmark 122,479 & \cmark 157,084 & \cmark 84,436 \\
\hline
\end{tabular}
\label{Non-IID_Severe_distribution}
\end{table*}

\subsection{Prototype generation via GMM}

To realise privacy-preserving knowledge sharing while mitigating class imbalance, every client succinctly encodes its local data through \emph{class-wise prototypes}.  Instead of transmitting raw samples or model gradients, each client fits a GMM to the feature vectors of every local class and shares only the perturbed component means.

\paragraph*{Local GMM fitting}
Let $\mathcal{C}_k$ denote the set of classes present on client~$k$ and  
$X_c=\{x_i\in\mathbb{R}^{d}\mid y_i=c\}$ the feature matrix of class $c\in\mathcal{C}_k$.
The class density is modelled as  
\begin{equation}
p(x\mid c)=\sum_{j=1}^{K_c}\pi_{j}^{(c)}
\;\mathcal{N}\!\bigl(x;\,\boldsymbol{\mu}_{j}^{(c)},\boldsymbol{\Sigma}_{j}^{(c)}\bigr),
\label{eq:gmm}
\end{equation}
where  

\begin{itemize}[leftmargin=1.5em]
    \item $K_c$ is the (data-driven) number of mixture components, selected via the Bayesian Information Criterion (BIC);
    \item $\pi_{j}^{(c)}>0$ is the weight of component~$j$ with $\sum_{j=1}^{K_c}\pi_{j}^{(c)}\!=\!1$;
    \item $\boldsymbol{\mu}_{j}^{(c)}\!\in\!\mathbb{R}^{d}$ and
          $\boldsymbol{\Sigma}_{j}^{(c)}\!\in\!\mathbb{R}^{d\times d}$ are, respectively, its mean (prototype) and covariance;
    \item $\mathcal{N}(\cdot;\boldsymbol{\mu},\boldsymbol{\Sigma})$ denotes the $d$-variate Gaussian probability density function.
\end{itemize}

\paragraph*{Prototype extraction and perturbation}
After training the GMM, the set  
$\{\boldsymbol{\mu}_{j}^{(c)}\}_{j=1}^{K_c}$ serves as the class prototypes.  
To protect privacy, each mean is obfuscated by additive noise:
\begin{equation}
\tilde{\boldsymbol{\mu}}_{j}^{(c)}=\boldsymbol{\mu}_{j}^{(c)}
+\boldsymbol{\epsilon},
\quad
\boldsymbol{\epsilon}\sim
\mathcal{N}\!\bigl(\mathbf{0},\,\sigma^{2}\mathbf{I}_{d}\bigr),
\label{eq:dp-noise}
\end{equation}
where  

\begin{itemize}[leftmargin=1.5em]
    \item $\sigma$ controls the noise amplitude (tuned empirically to balance privacy and utility);
    \item $\mathbf{I}_{d}$ is the $d\times d$ identity matrix;
    \item $\tilde{\boldsymbol{\mu}}_{j}^{(c)}$ is the privacy-preserving prototype that will be uploaded.
\end{itemize}

\paragraph*{Conditional exchange}
The prototype-sharing routine is \emph{one-off and adaptive}:  
the server monitors the global accuracy during the first five rounds and triggers an exchange only if the mean accuracy drops below a preset threshold (97\,\% in our experiments). However, these hyper-parameters are configurable in real-world deployments depending on the dataset characteristics, application requirements, and the convergence behavior of the federated system.

\paragraph*{Rationale}
Sharing perturbed GMM means fulfils three goals simultaneously:  
(1)~it keeps raw data local,  
(2)~it conveys essential class statistics to the federation, and  
(3)~it injects privacy noise that thwarts sample reconstruction.  
Consequently, each client gains statistical cues about under-represented or unseen classes, enabling more balanced updates when local data are skewed.

\subsection{Prototype aggregation and redistribution}
To enable scalable and privacy-aware knowledge transfer, we introduce a centralized prototype aggregation mechanism that consolidates noisy client-side class representations. Once clients transmit their class-wise prototypes during the designated exchange round, the server initiates a centralized aggregation process. For each class $c$ observed across the federation, the server collects the corresponding set of perturbed prototype vectors $\{\tilde{\mu}_j^{(c)}\}$ from all contributing clients. These vectors are then clustered using the MiniBatch K-Means algorithm, which is computationally efficient and well-suited for processing large volumes of prototype data with minimal overhead. The goal is to derive a compact yet representative set of global prototypes $\{\hat{\mu}_l^{(c)}\}_{l=1}^{L_c}$ for each class, where $L_c$ denotes the number of clusters selected heuristically based on the number of received prototypes and the dimensionality of the feature space.

This aggregated set of global prototypes captures a more comprehensive view of class-specific distributions across the federated system. In the subsequent communication round, the server redistributes these refined prototypes to all participating clients. By integrating these global class representations into their local training pipelines, clients obtain statistical cues for classes that may not be present in their local datasets. This enhances the learning process and promotes better generalization across heterogeneous data. 

Importantly, this redistribution strategy enables clients with skewed or disjoint data to benefit from the broader class diversity of the federation without violating data privacy constraints, thereby contributing to a more robust and balanced global model.

\subsection{Data augmentation with SMOTE}
To further mitigate local class imbalance and enrich the training corpus, clients apply the Synthetic Minority Oversampling Technique (SMOTE) to the received global prototypes. Upon receiving the set $\{\hat{\mu}_l^{(c)}\}$ for underrepresented class $c$, each client generates additional synthetic samples by interpolating between randomly selected prototype pairs in the feature space. The number of synthetic instances is controlled to achieve approximately a 10\% increase in the training set size relative to the number of received prototypes.

Formally, let $\hat{\mu}_a^{(c)}, \hat{\mu}_b^{(c)} \in \mathbb{R}^d$ be two randomly chosen prototype vectors corresponding to class $c$, where $d$ is the feature space dimension. A synthetic sample $x_{\text{syn}} \in \mathbb{R}^d$ is then generated as follows:
\begin{equation}
x_{\text{syn}} = \hat{\mu}_a^{(c)} + \lambda \cdot (\hat{\mu}_b^{(c)} - \hat{\mu}_a^{(c)}), \quad \lambda \sim \mathcal{U}(0,1),
\end{equation}
where $\lambda$ is sampled from a continuous uniform distribution $\mathcal{U}(0,1)$. This linear interpolation preserves the geometric coherence of the class distribution while introducing controlled variability, thereby enhancing the representation of minority classes during local training.

The combination of prototype redistribution and SMOTE-based augmentation equips each client with a synthesized and more balanced perspective of the data space. This approach indirectly enhances model robustness—particularly in the presence of highly skewed or disjoint local datasets—without ever transmitting raw data samples or client-specific features.

\vspace{1em}

Together, these components form the backbone of the proposed FedP3E framework, offering a scalable and privacy-preserving mechanism for indirect knowledge transfer across non-IID and imbalanced IoT malware datasets within cross-silo FL environments.

\section{Experiments and Analysis}
\label{Experiments and Analysis}

\subsection{Experimental setup}
\label{sec:exp_setup}

This subsection outlines the data–preprocessing pipeline, the hardware/software stack, and the FL configuration. The full set of training hyper-parameters is listed in Table~\ref{tab:mlp_fl_hparams}.

\subsubsection{Preprocessing}
Each client loads its local CSV partition, scales all features to $[0,1]$ with Min–Max normalisation, and performs an 80/20 stratified split to create training and test subsets. No further feature engineering or dimensionality reduction is applied, resulting in an input dimension of 115 features.

\subsubsection{Hardware and software environment}
Experiments run on a Linux workstation equipped with two NVIDIA GeForce RTX 4090 GPUs (24 GB each) and CUDA 12.2.  Implementation uses Python 3.11.5, TensorFlow 2.15 (Keras API), and the Flower FL framework v1.15.2.

\subsubsection{FL Configuration}
We follow a cross-silo FL setup consisting of three high-capacity clients and a central server. All clients participate in every round, training the model locally for 15 epochs per round across 20 communication rounds. After the initial five rounds, if the global model’s average accuracy falls below 97\%, a one-time prototype exchange is triggered at round 6. In this step, each client transmits class-wise GMM centroids with added Gaussian noise to preserve data privacy, which the server reclusters the received centroids via MiniBatch K-Means and broadcasts the resulting global prototypes.  
Clients then apply SMOTE to augment minority classes by 10 \% (relative to the original training size) and continue training.

\begin{table}[t]
  \caption{P3E Federated-Learning Hyper-Parameters}
  \label{tab:mlp_fl_hparams}
  \centering
  \footnotesize
  \setlength{\tabcolsep}{3.5pt}
  \begin{tabular}{@{}p{3.9cm}p{4.7cm}@{}}
    \toprule
    \textbf{Component / Setting} & \textbf{Configuration} \\
    \midrule
    \multicolumn{2}{@{}l}{\textbf{Neural Network Architecture}} \\
    Input layer         & 115 numeric features \\
    Dense layer 1       & 128 units, ReLU, L\textsubscript{2}= 0.001 \\
    Batch normalization & After Dense\,1 \\
    Dropout             & $p=0.5$ after Dense\,1 \\
    Dense layer 2       & 64 units, ReLU, L\textsubscript{2}= 0.001 \\
    Batch normalization & After Dense\,2 \\
    Output layer        & 3 units, Softmax \\
    \midrule
    \multicolumn{2}{@{}l}{\textbf{Local Training Settings}} \\
    Optimizer           & Adam, learning rate = 0.0001 \\
    Loss function       & Sparse categorical cross‐entropy \\
    Epochs per round    & 15 \\
    Batch size          & 32 \\
    \midrule
    \multicolumn{2}{@{}l}{\textbf{Federated Learning Setup}} \\
    FL strategy         & FedAvg (full participation) \\
    Number of rounds    & 20 \\
    Prototype exchange  & Triggered if mean accuracy (rounds 1–5) $<0.97$ \\
    Gaussian noise      & Added to prototypes, $\sigma = 0.01$ \\
    \midrule
    \textbf{Total trainable parameters} & 23,683 \\ 
    \bottomrule
  \end{tabular}
\end{table}

\subsection{Baseline methods}

To benchmark the performance of the proposed P3E framework, we conduct comparative experiments under both IID and non-IID data distributions using two widely adopted FL algorithms: FedAvg and FedProx. These baseline methods are configured with the same training-level hyperparameters as FedP3E, excluding the prototype exchange mechanism and the incorporation of Gaussian noise. For FedProx, we evaluate its performance under varying values of the proximal term $\mu$, specifically $\mu \in \{0.1, 0.3, 1.0\}$, to assess its sensitivity to different regularization strengths.

\begin{table*}[!t]
\caption{End-of-training performance of all Federated Learning methods under IID and non-IID data-distribution scenarios}
\label{tab:perf_all_scenarios}
\centering
\small        
\setlength{\tabcolsep}{4pt}
\renewcommand{\arraystretch}{1.1}
\begin{tabular}{lrrrrrr}
\toprule
\textbf{Method} & \textbf{Accuracy} & \textbf{Precision} & \textbf{Recall} & \textbf{F1-score} & \textbf{Loss} & \textbf{Training time (s)}\\
\midrule
\multicolumn{7}{c}{\textbf{Scenario 1: IID}}\\
\midrule
FedP3E                  & 0.9971 & 0.997 & 0.997 & 0.997 & 0.0168 & 17,346.7\\
FedAvg                  & 0.9974 & 0.997 & 0.997 & 0.997 & 0.0214 & 17,171.2\\
FedProx ($\mu=0.1$)     & 0.9890 & 0.989 & 0.989 & 0.989 & 0.0423 & 17,186.4\\
FedProx ($\mu=0.3$)     & 0.9059 & 0.906 & 0.906 & 0.906 & 0.4927 & 17,414.6\\
FedProx ($\mu=1.0$)     & 0.1267 & 0.127 & 0.127 & 0.127 & 1.1750 & 16,937\\
\midrule
\multicolumn{7}{c}{\textbf{Scenario 2 - Case 1: Light non-IID}}\\
\midrule
FedP3E                  & 0.9957 & 0.996 & 0.995 & 0.996 & 0.0199 & 12,338.3\\
FedAvg                  & 0.9379 & 0.940 & 0.935 & 0.938 & 0.3870 &  7,553\\
FedProx ($\mu=0.1$)     & 0.9278 & 0.930 & 0.925 & 0.928 & 0.3486 &  7,599.7\\
FedProx ($\mu=0.3$)     & 0.9610 & 0.961 & 0.960 & 0.961 & 0.1443 &  7,493.2\\
FedProx ($\mu=1.0$)     & 0.3328 & 0.330 & 0.330 & 0.331 & 1.1667 &  7,571.5\\
\midrule
\multicolumn{7}{c}{\textbf{Scenario 2 - Case 2: Moderate non-IID}}\\
\midrule
FedP3E                  & 0.9940 & 0.994 & 0.994 & 0.994 & 0.0283 & 12,040.9\\
FedAvg                  & 0.9196 & 0.920 & 0.918 & 0.919 & 0.3921 &  8,052.6\\
FedProx ($\mu=0.1$)     & 0.9247 & 0.926 & 0.923 & 0.925 & 0.3407 &  7,949.9\\
FedProx ($\mu=0.3$)     & 0.9402 & 0.941 & 0.940 & 0.940 & 0.2167 &  8,103.2\\
FedProx ($\mu=1.0$)     & 0.5911 & 0.590 & 0.590 & 0.590 & 1.1379 &  7,971.4\\
\midrule
\multicolumn{7}{c}{\textbf{Scenario 2 - Case 3: Severe non-IID}}\\
\midrule
FedP3E                  & 0.9511 & 0.951 & 0.950 & 0.951 & 0.7251 &  8\,663.3\\
FedAvg                  & 0.4939 & 0.494 & 0.494 & 0.494 & 1.3466 &  5\,788\\
FedProx ($\mu=0.1$)     & 0.4939 & 0.494 & 0.494 & 0.494 & 1.3556 &  5\,785.5\\
FedProx ($\mu=0.3$)     & 0.4939 & 0.494 & 0.494 & 0.494 & 1.3637 &  5\,791.1\\
FedProx ($\mu=1.0$)     & 0.0485 & 0.049 & 0.048 & 0.048 & 1.2224 &  5\,707.7\\
\bottomrule
\end{tabular}
\vspace{-1ex}
\end{table*}

\subsection{Results and analysis}
This subsection presents a comprehensive evaluation of the proposed FedP3E framework across various data distribution scenarios. We compare its performance with FedAvg and FedProx using standard classification metrics, including accuracy, precision, recall, F1-score, and communication efficiency. Table~\ref{tab:perf_all_scenarios} summarizes the end-of-training performance of all FL methods under both IID and non-IID conditions.

\subsubsection{IID scenario}

\begin{figure}[!t]
\centering
\includegraphics[width=\columnwidth]{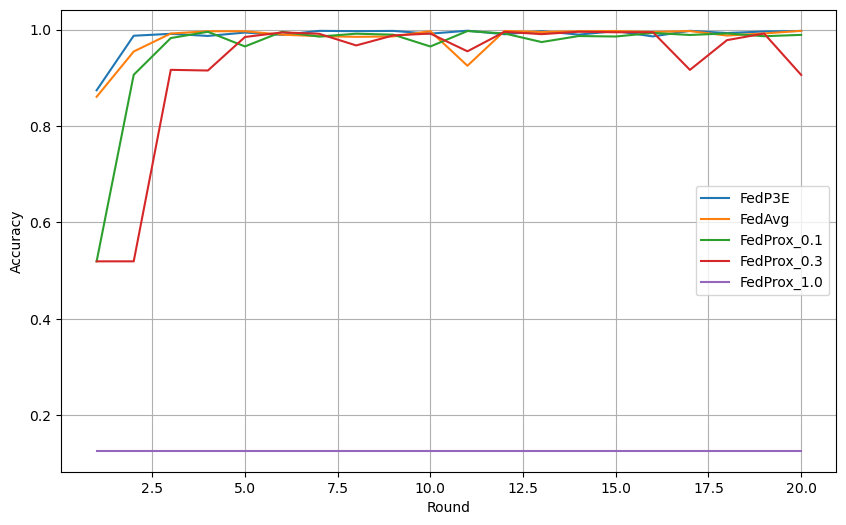}
\caption{Comparison of accuracy across 20 rounds under the IID data distribution scenario for FedP3E, FedAvg, and FedProx with varying $\mu$ values.}
\label{acc_iid}
\end{figure}

\begin{figure}[!t]
\centering
\includegraphics[width=\columnwidth]{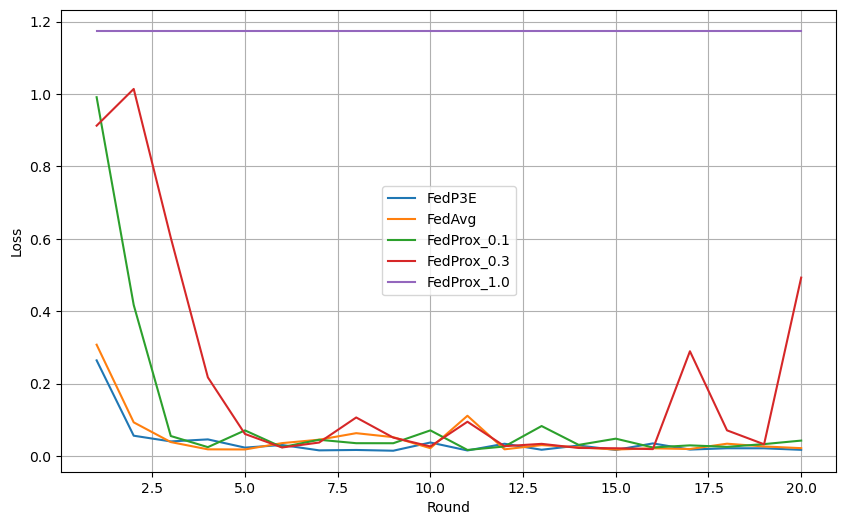}
\caption{Evolution of training loss across 20 rounds under the IID data distribution scenario for FedP3E, FedAvg, and FedProx with varying $\mu$ values.}
\label{loss_iid}
\end{figure}

In the IID scenario, both FedAvg and FedP3E rapidly exceed 99.5\% accuracy within the first 10 communication rounds, with FedP3E attaining the lowest final loss ($0.0168$) and a slightly smoother learning curve. In contrast, FedProx with $\mu=0.1$ trails slightly and plateaus at 98.9\% accuracy. As the regularization strength increases to $\mu=0.3$, convergence further degrades, and the final accuracy drops to 90.6\%. With $\mu=1.0$, FedProx struggles to learn altogether. These trends underscore that heavy proximal regularization is unnecessary in balanced data environments. The detailed accuracy and loss trends are shown in Fig.~\ref{acc_iid} and~\ref{loss_iid}, respectively.

\begin{tcolorbox}[colback=gray!10, colframe=gray!80, sharp corners, boxrule=0.4pt]
\textbf{Takeaway:} In balanced IID settings, FedP3E operates comparably to FedAvg without activating its prototype-exchange mechanism, which remains dormant unless a performance drop is observed. This allows the framework to maintain high performance with minimal communication overhead. In contrast, FedProx exhibits slower convergence and reduced final accuracy, indicating that proximal regularization offers limited value in such balanced environments.
\end{tcolorbox}

\subsubsection{Light non-IID}

\begin{figure}[!t]
\centering
\includegraphics[width=\columnwidth]{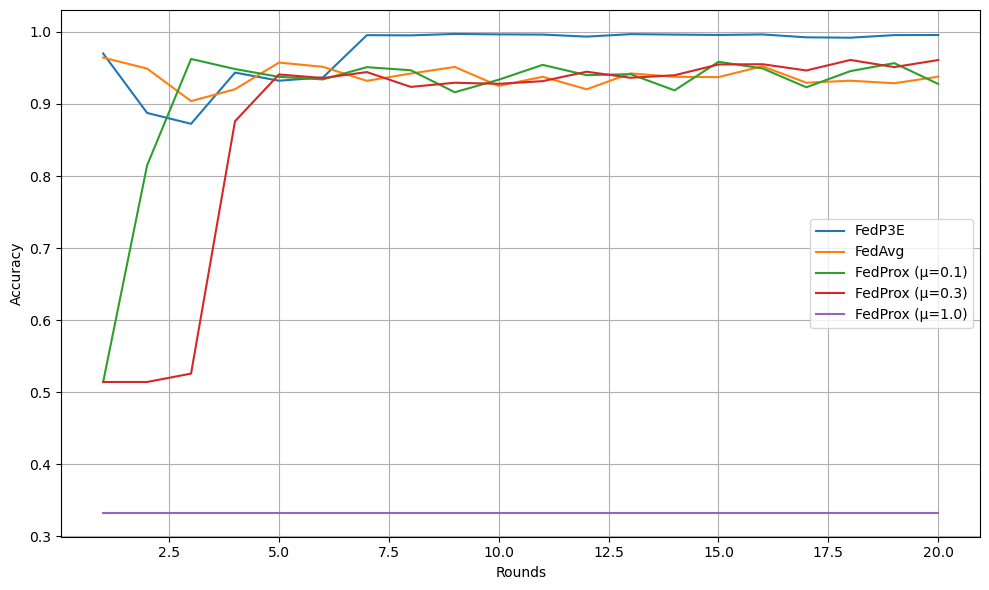}
\caption{Comparison of accuracy across 20 rounds under the light non-IID data distribution scenario for FedP3E, FedAvg, and FedProx with varying $\mu$ values.}
\label{acc_light}
\end{figure}

In the light non-IID setting, label distribution skew is mild: while all clients observe samples from both malware families, the frequency of individual variants differs across them. This setup reflects realistic deployments, where certain IoT environments experience uneven exposure to specific attack types. As shown in Fig.~\ref{acc_light}, FedP3E experiences a slight initial drop in accuracy—due to the deliberate class imbalance—but once the average global accuracy at round 5 falls below the 97\% threshold, the one-time prototype exchange is triggered. This immediately improves performance, with accuracy rising from 93.3\% (round 6) to 99.5\% (round 7), and remaining above 99\% thereafter.

\begin{figure}[!t]
\centering
\includegraphics[width=\columnwidth]{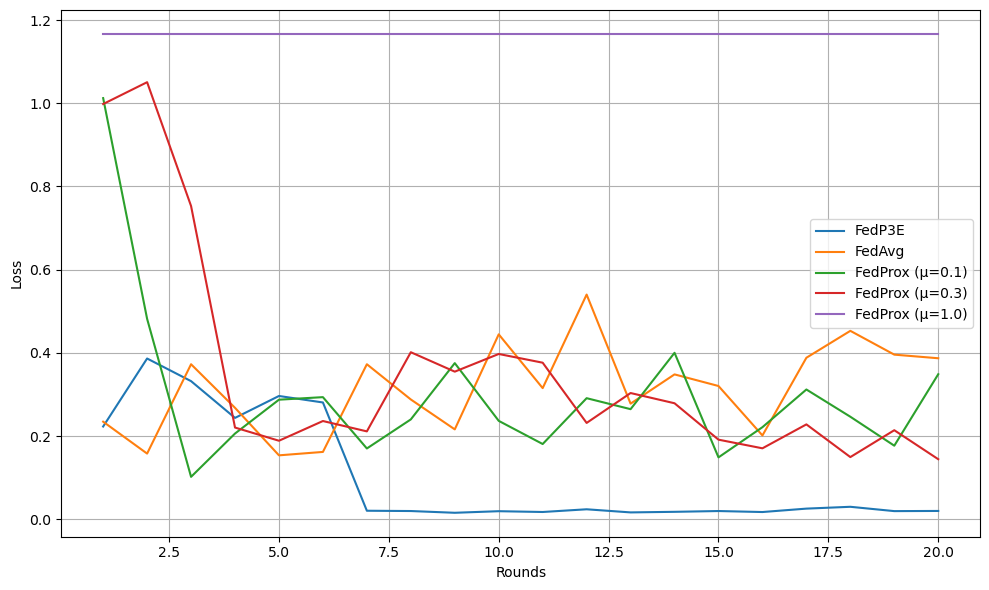}
\caption{Evolution of training loss across 20 rounds under the light non-IID data distribution scenario for FedP3E, FedAvg, and FedProx with varying $\mu$ values.}
\label{loss_light}
\end{figure}

Fig.~\ref{loss_light} further illustrates that this adjustment also leads to a significant drop in cross-entropy loss, reinforcing the stabilizing effect of prototype redistribution. In contrast, FedAvg and FedProx (with $\mu=0.1$ and $0.3$) converge more slowly and plateau between 93–96\% accuracy, while FedProx with strong regularization ($\mu=1.0$) fails to learn, stalling at just 33.28\% accuracy. By round 20, FedP3E achieves an F1-score of 99.6\% and a loss of 0.020—improving by approximately 3.5 percentage points in F1-score and 86\% in loss relative to the next-best baseline (FedProx, $\mu=0.3$). The performance gain is most pronounced on minority variants, with recall improving by about 6 percentage points compared to FedAvg, indicating that prototype redistribution effectively compensates for mild class imbalance.

\begin{tcolorbox}[colback=gray!10, colframe=gray!80, sharp corners, boxrule=0.4pt]
\textbf{Takeaway:}
Under light heterogeneity, a one-time, privacy-preserving prototype exchange is sufficient to close the statistical gap between clients, enabling near-optimal global performance without data exposure or heavy communication. These results illustrate the practicality of FedP3E in scenarios where mild non-IID label imbalance exists.
\end{tcolorbox}

\subsubsection{Moderate non-IID}

In the moderate non-IID setting, several malware variants are entirely absent from one or more clients. This heavier statistical skew (Table~\ref{Non-IID_Moderate_distribution}) emulates scenarios where clients operate in isolation or lack exposure to specific threat categories—for example, \textit{udp plain} (Mirai) appears only at Client~2, while \textit{combo} (Gafgyt) is limited to Client~1.

\begin{figure}[!t]
\centering
\includegraphics[width=\columnwidth]{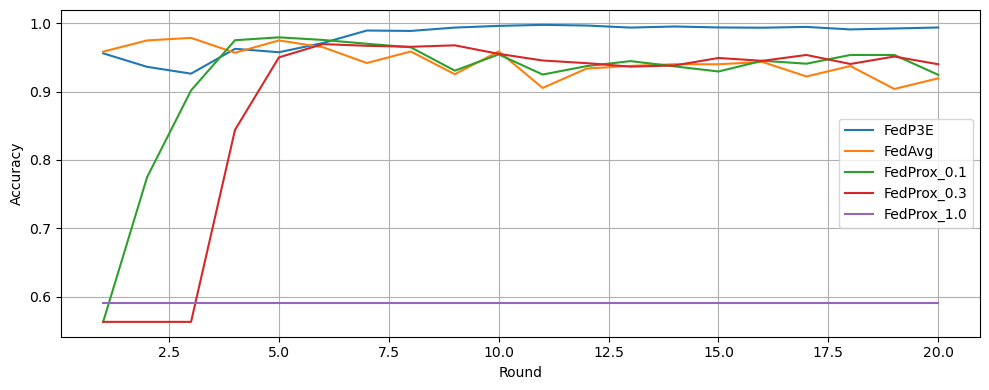}
\caption{Comparison of accuracy across 20 rounds under the moderate non-IID data distribution scenario for FedP3E, FedAvg, and FedProx with varying $\mu$ values.}
\label{acc_moderate}
\end{figure}

\begin{figure}[!t]
\centering
\includegraphics[width=\columnwidth]{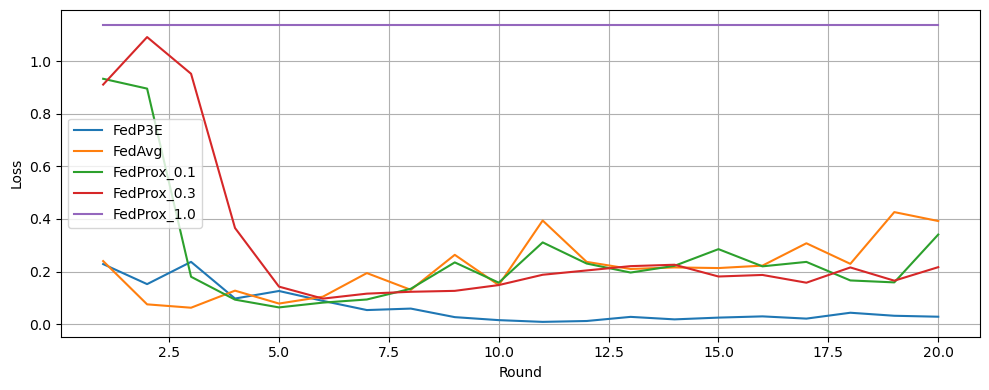}
\caption{Evolution of training loss across 20 rounds under the moderate non-IID data distribution scenario for FedP3E, FedAvg, and FedProx with varying $\mu$ values.}
\label{loss_moderate}
\end{figure}

FedP3E demonstrates strong resilience to this form of heterogeneity. As illustrated in Fig.\ref{acc_moderate}, it steadily improves throughout the training process, achieving an accuracy and F1-score of 99.4\% by round 20, along with the lowest final loss (0.028) as shown in Fig.~\ref{loss_moderate}. These gains are supported by a single round of prototype redistribution, which provides indirect exposure to unseen classes and facilitates inter-client knowledge transfer.

Among baseline methods, FedProx with $\mu = 0.3$ delivers the closest performance, plateauing near 96.1\% accuracy, while FedAvg struggles with fluctuating accuracy and higher loss values. FedProx with $\mu = 1.0$ fails to converge entirely, indicating that aggressive regularization may be counterproductive in class imbalanced environments.

Importantly, FedP3E yields notable improvements in minority-class recall—achieving gains of approximately 7 percentage points over FedAvg and around 5 points over FedProx ($\mu = 0.3$). This highlights the effectiveness of targeted prototype exchange in moderating distribution skew while preserving both communication efficiency and client data privacy.

\begin{tcolorbox}[colback=gray!10, colframe=gray!80, sharp corners, boxrule=0.4pt]
\textbf{Takeaway:}
Under moderate non-IID conditions, FedP3E demonstrates strong generalization by using a one-time prototype redistribution to ensure all clients benefit from class information missing in their local datasets. Its consistent performance and improved recall on minority variants outperform all baselines, including the best FedProx variant ($\mu = 0.3$), highlighting its effectiveness in handling statistical heterogeneity without compromising privacy or communication efficiency.
\end{tcolorbox}

\subsubsection{Severe non-IID}

The severe non-IID scenario represents the most challenging distribution setting, where each client holds data from a mutually exclusive class subset, resulting in complete label disjointness across the federation. Specifically, Client~1 contains only benign samples, Client~2 processes Gafgyt-infected traffic, and Client~3 receives only Mirai instances. This setting emulates highly siloed deployments—such as separate enterprise departments, segmented sensor networks, or distinct security zones—where operational roles and device types lead to drastically different threat exposures and data profiles.

\begin{figure}[!t]
\centering
\includegraphics[width=\columnwidth]{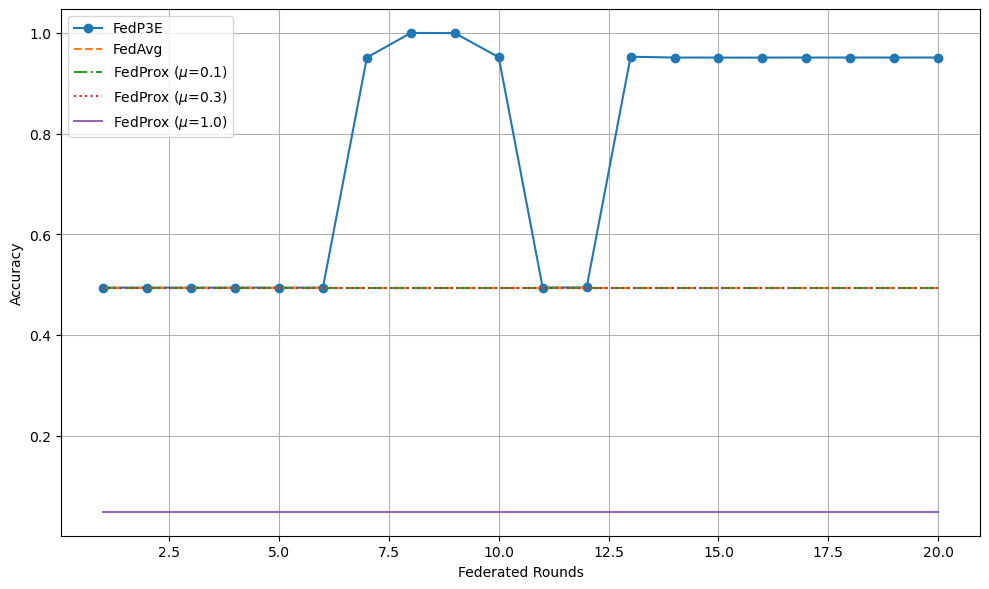}
\caption{Comparison of accuracy across 20 rounds under the severe non-IID data distribution scenario for FedP3E, FedAvg, and FedProx with varying $\mu$ values.}
\label{acc_severe}
\end{figure}

Under this setting, baseline FL methods struggle to generalize due to the absence of overlapping label spaces across clients. As shown in Fig.~\ref{acc_severe}, FedAvg fails to learn meaningful patterns, with global accuracy plateauing around 50\% throughout. Similarly, FedProx across all tested $\mu$ values shows poor convergence, as clients lack exposure to any class diversity and the aggregation process struggles to align incompatible local models.

In contrast, FedP3E exhibits remarkable resilience. Despite the lack of overlapping label spaces, its one-time exchange of class-wise prototypes allows each client to gain semantically enriched representations of unseen categories. Following this exchange, SMOTE is applied locally to synthesize a small number (10\%) of new training samples for missing classes based on the received prototypes. This enhancement empowers each client to update its model with representative features of all classes, enabling the server to aggregate a more coherent and generalizable global model.

\begin{figure}[!t]
\centering
\includegraphics[width=\columnwidth]{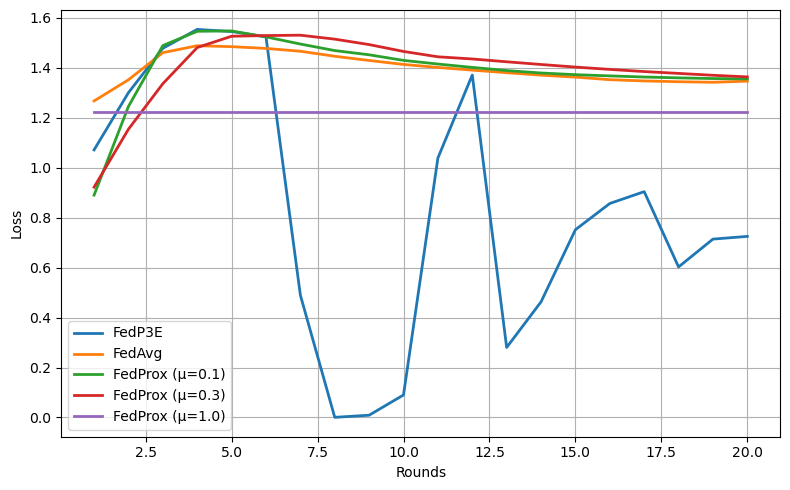}
\caption{Evolution of training loss across 20 rounds under the severe non-IID data distribution scenario for FedP3E, FedAvg, and FedProx with varying $\mu$ values.}
\label{loss_severe}
\end{figure}

As illustrated in Fig.~\ref{loss_severe}, FedP3E significantly reduces loss after the prototype phase and maintains stable convergence thereafter. While the approach incurs a modest increase in communication due to prototype sharing, the performance gain—in both accuracy and convergence—is substantial, making FedP3E a practical solution for real-world federated deployments characterized by severe non-IID.

\begin{tcolorbox}[colback=gray!10, colframe=gray!80, sharp corners, boxrule=0.4pt]
\textbf{Takeaway:}
FedP3E effectively addresses the limitations of baseline FL algorithms in severe non-IID settings by introducing cross-client semantic context through prototype sharing and local SMOTE-based augmentation. This strategy enables robust convergence even in completely disjoint class distributions, where client datasets are inherently isolated and unbalanced. In contrast, baseline methods such as FedAvg and FedProx fail to generalize under such conditions, as they lack mechanisms to compensate for the absence of shared label space across clients.
\end{tcolorbox}

\subsubsection{FedAvg vs. FedProx: Comparative analysis across $\mu$ variants}

FedAvg operates by averaging client model updates without regularization, while FedProx introduces a proximal term governed by the parameter $\mu$ to constrain local updates from drifting too far from the global model. This design aims to stabilize training in non-IID scenarios by aligning local objectives more closely with the global model.

However, our results reveal significant limitations in FedProx’s performance across different levels of data heterogeneity. Even under balanced IID conditions—where clients share identical class distributions—FedProx with $\mu = 1.0$ performs poorly, reaching only 12\% accuracy. This drastic underperformance results from overly restrictive regularization, which severely limits local model updates and prevents the model from learning even in well-aligned data settings. Such behavior suggests that strong proximal constraints can be counterproductive in scenarios where client data is already consistent and does not require additional stabilization. In the light non-IID scenario, FedProx with $\mu = 0.3$ achieves slightly better final accuracy than both FedAvg and FedProx with $\mu = 0.1$, while maintaining a comparable convergence rate. In the moderate non-IID setting, FedProx with $\mu = 0.1$ and $0.3$ outperform FedAvg in terms of accuracy, although all three approaches experience instability across rounds. This is primarily due to the absence of specific variant-level on certain clients. Without a mechanism to compensate for these missing variants, the locally-trained models struggle with consistent predictions on under-represented categories. In the severe non-IID case, where each client has access to only a single class, all FedProx variants exhibit a dramatic collapse in performance. The proximal term, which is intended to keep local models aligned with the global model, ends up restricting their ability to learn from the limited data available. At the same time, it provides no mechanism for handling unseen classes, which are entirely absent from local datasets. 

\begin{tcolorbox}[colback=gray!10, colframe=gray!80, sharp corners, boxrule=0.4pt]
\textbf{Takeaway:}
FedProx's effectiveness depends heavily on the data distribution and the choice of $\mu$. Its performance drops significantly in both balanced and skewed settings when regularization is too aggressive. In federated environments with high class imbalance or disjoint label spaces, stronger strategies like FedP3E—which actively enhance each client’s view of the global class distribution—are required to ensure stable training and generalization.
\end{tcolorbox}

\subsubsection{Communication cost}

To evaluate the communication efficiency of FedP3E, we quantify the transmission overhead associated with its prototype exchange mechanism relative to standard FL protocols. Let the communication payloads be defined as:
\[
\mathbf{W} \in \mathbb{R}^{d_w}, \qquad
\mathbf{P}_i = \left\{ \boldsymbol{\mu}_{k,j} \right\}_{%
    \substack{k = 1, \dots, C\\ j = 1, \dots, m_k}}, \quad
    \boldsymbol{\mu}_{k,j} \in \mathbb{R}^{d_x},
\]
where:
\begin{itemize}[leftmargin=1.6em]
    \item $d_w = 23{,}683$ — total number of trainable model parameters,
    \item $C$ — number of classes observed locally by client~$i$,
    \item $m_k$ — number of GMM prototypes per class~$k$,
    \item $d_x = 115$ — dimensionality of the input feature space,
    \item $\lvert \mathbf{W} \rvert = d_w$ — size of the model update in floats,
    \item $\lvert \mathbf{P}_i \rvert = \sum_{k=1}^{C} m_k \cdot d_x$ — total size of prototype vectors uploaded by client~$i$.
\end{itemize}

\paragraph*{Regular FL Communication}
In each FL round, clients upload their local model and receive the global model:
\[
\text{Upload: } \mathbf{W}, \quad
\text{Download: } \mathbf{W}^{(\text{global})} \quad \Rightarrow \quad 
\text{Total: } 2 \cdot \lvert \mathbf{W} \rvert \text{ floats}.
\]

\paragraph*{Prototype Exchange Overhead}
If the global model fails to reach a predefined accuracy threshold (97\%) after five communication rounds, a one-time prototype exchange is initiated. Each client uploads class-wise perturbed GMM centroids $\mathbf{P}_i$ and receives a global prototype set $\widehat{\mathbf{P}}$ aggregated and reclustered by the server:
\[
\text{Upload: } \mathbf{P}_i, \quad
\text{Download: } \widehat{\mathbf{P}}.
\]

\paragraph*{Upload Size}
Assuming each client observes $C = 3$ classes and generates $m_k = 3$ prototypes per class, the total prototype upload size becomes:
\[
\lvert \mathbf{P}_i \rvert = C \cdot m_k \cdot d_x = 3 \cdot 3 \cdot 115 = 1035 \text{ floats}.
\]
Relative to the full model, this represents an upload cost of:
\[
\frac{\lvert \mathbf{P}_i \rvert}{\lvert \mathbf{W} \rvert} = \frac{1035}{23{,}683} \approx 4.37\%.
\]

\paragraph*{Download Size}
Following aggregation, the server sends back $\widehat{\mathbf{P}}$ consisting of $m_k' = 4$ prototypes per class for $C_{\text{global}} = 3$ global classes:
\[
\lvert \widehat{\mathbf{P}} \rvert = C_{\text{global}} \cdot m_k' \cdot d_x = 3 \cdot 4 \cdot 115 = 1380 \text{ floats},
\]
yielding a relative download cost of:
\[
\frac{\lvert \widehat{\mathbf{P}} \rvert}{\lvert \mathbf{W} \rvert} = \frac{1380}{23{,}683} \approx 5.83\%.
\]

\paragraph*{Total Prototype Communication.}
The total cost of the one-time prototype exchange, accounting for both upload and download, is:
\[
\frac{\lvert \mathbf{P}_i \rvert + \lvert \widehat{\mathbf{P}} \rvert}{\lvert \mathbf{W} \rvert} = \frac{1035 + 1380}{23{,}683} \approx 9.58\%.
\]

\begin{tcolorbox}[colback=gray!10, colframe=gray!80, sharp corners, boxrule=0.4pt]
\textbf{Takeaway:}
FedP3E introduces a one-time prototype exchange mechanism that incurs less than 10\% of the communication volume of a full model round. This lightweight overhead enables statistically enriched updates under non-IID conditions while preserving the communication efficiency of standard FL protocols in all other rounds.
\end{tcolorbox}

\subsubsection{Training overhead}
\begin{figure}[!t]
\centering
\includegraphics[width=\columnwidth]{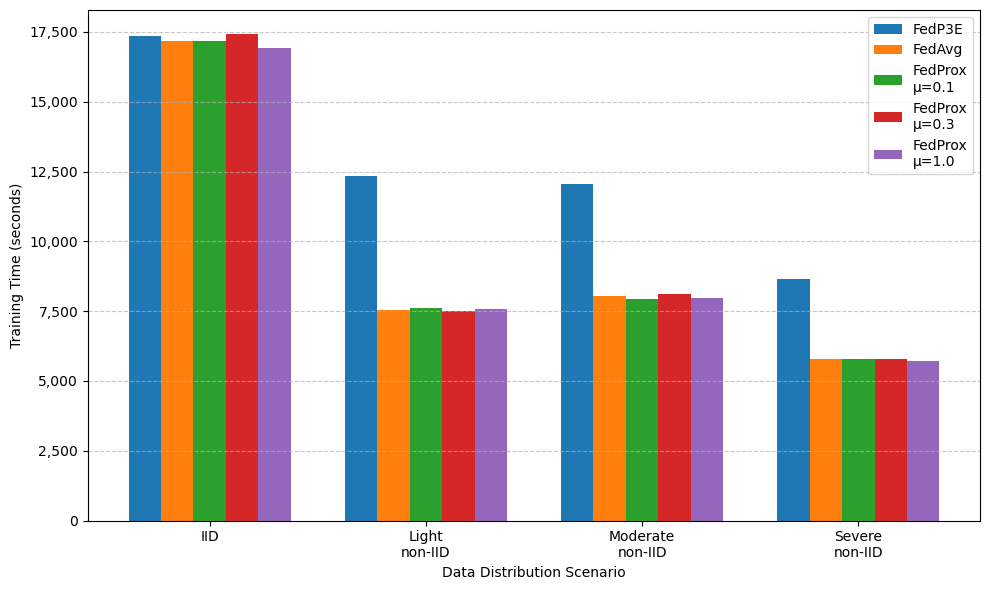}
\caption{Training time comparison across IID and non-IID scenarios.}
\label{training_time}
\end{figure}

In addition to communication efficiency and model accuracy, we evaluate the training overhead of FedP3E across varying degrees of data heterogeneity. All methods were executed on identical hardware to ensure fair comparison, and Fig.~\ref{training_time} presents the total training time under IID, light, moderate, and severe non-IID scenarios.

Under the IID scenario, all approaches yield nearly identical training durations, indicating that FedP3E introduces no computational penalty when prototype exchange remains inactive. This confirms its ability to remain lightweight under balanced conditions.

However, as data heterogeneity increases, FedP3E incurs a higher training cost compared to FedAvg and FedProx. This is attributed to the additional steps of prototype generation, centralized aggregation, and SMOTE-based data augmentation triggered by non-IID conditions. While this overhead is non-negligible, it reflects a deliberate trade-off: enhanced generalization and robustness in challenging settings, achieved without any repeated exchange or raw data sharing.

\begin{tcolorbox}[colback=gray!10, colframe=gray!80, sharp corners, boxrule=0.4pt]
\textbf{Takeaway:}
FedP3E remains lightweight in IID settings and activates its additional components only when performance falls below a threshold. Although this introduces moderate computational overhead in skewed scenarios, the improved robustness justifies the cost in real-world deployments.
\end{tcolorbox}

\subsection{Comparison with existing federated learning methods on N-BaIoT}
To further highlight the performance of FedP3E, Table~\ref{fl_nbaioit_comparison} provides a comparative analysis against existing FL-based methods evaluated on the N-BaIoT dataset. Most prior studies focus on standard IID or mildly skewed non-IID scenarios and do not explicitly account for fully disjoint class distributions. As a result, their effectiveness diminishes as data heterogeneity increases. In contrast, FedP3E is explicitly designed to operate across the full spectrum of data distributions, including extreme cases where each client possesses samples from only one class. It achieves 99.71\% accuracy in the IID setting, 99.57\% under light non-IID, 99.40\% in moderate non-IID, and maintains a strong 95.11\% accuracy under severe non-IID conditions. This consistent performance across increasing degrees of heterogeneity underscores FedP3E’s adaptability and robustness, making it a practical and scalable solution for federated IoT malware detection in highly diverse and siloed environments.

\begin{table}[!t]
\caption{Comparison of Proposed FedP3E with Existing Federated Learning-based Methods on N-BaIoT Dataset}
\label{fl_nbaioit_comparison}
\centering
\begin{tabular}{|p{1cm}|p{1.2cm}|p{1.5cm}|p{2.9cm}|}
\hline
\textbf{Ref.} & \textbf{IID OR}  \newline \textbf{Non-IID?} & \textbf{Disjoint} \newline \textbf{Class} \newline \textbf{Considered?} & \textbf{Accuracy (\%)} \\
\hline
\cite{rey2022federated} & Non-IID & \xmark & Supervised: 99.42--99.92 \newline Unsupervised: $\approx$97.4 \\
\hline
\cite{zhou2023federated} & IID & \xmark & 99.63 \\
\hline
\cite{do2024horizontal} & Non-IID & \xmark & 90.00 \\
\hline
\cite{gupta2025distributed} & Non-IID & \xmark & 89.90 \\
\hline
\textbf{FedP3E}& Both & \cmark & IID: 99.71 \newline Non-IID: 95.11--99.57 \\
\hline
\end{tabular}
\end{table}

\section{Conclusion and Future Work}
\label{Conclusion and Future Work}
This study proposed FedP3E, a prototype-based FL framework tailored to address the challenges of non-IID data distributions in IoT malware detection. Unlike conventional aggregation methods such as FedAvg and FedProx, FedP3E leverages class-wise prototype exchange to bridge the data heterogeneity gap across clients without compromising data privacy.

Extensive experiments were conducted across four data distribution scenarios—IID, light, moderate, and severe non-IID. The results demonstrate that FedP3E consistently achieves faster convergence, higher accuracy, and improved robustness compared to existing baselines, particularly under severe non-IID conditions. Notably, while FedProx is designed for non-IID settings, our experiments revealed its limitations in more skewed scenarios, where it failed to recover performance.

The findings underscore the effectiveness and scalability of prototype-based communication in real-world cross-silo FL settings where data imbalance and heterogeneity are prevalent. FedP3E offers a lightweight and privacy-preserving mechanism that enhances collaborative learning among distributed IoT devices, contributing to more secure and adaptive malware detection.

Looking ahead, we plan to deploy FedP3E in real-world on-device FL settings to assess its efficiency on lightweight IoT devices. This will enable us to evaluate the framework under more stringent constraints, including limited computational resources, constrained memory and communication bandwidth, and a vast number of dynamically selected heterogeneous devices with varying hardware capabilities.
In parallel, we intend to explore more complex forms of data heterogeneity, particularly feature-space skew, where clients have access to different subsets of input features. This presents new challenges for collaborative learning, requiring novel strategies for model alignment and robust generalization across diverse feature representations.

\section*{Acknowledgments}
This should be a simple paragraph before the References to thank those individuals and institutions who have supported your work on this article.

\bibliographystyle{IEEEtran}
\bibliography{bibfile_new}
\newpage

\vfill

\end{document}